\documentclass[12pt,a4paper]{article}

\usepackage{amsmath,amssymb,amsthm,epsfig}
\usepackage{rotating}

\newcommand{\C}{\mathbb{C}}

\newcommand{\e}{{\rm e\,}}
\newcommand{\1}{\mbox{1\hspace{-.7ex}I}}
\newcommand{\one}{\mbox{1\hspace{-.7ex}I}}
\newcommand{\tr}{{\rm tr}}
\newcommand{\Tr}{{\rm tr_{D,C}}\,}

\newcommand{\nn}{\nonumber}
\newcommand{\zu}{\zeta\,}
\newcommand{\zd}{{\tilde \zeta}\,}

\def\beal#1\eeal{\begin{align}#1\end{align}}

\newcommand{\cC}{{\cal C}}

\newcommand{\no}{\noindent}
\def\beal#1\eeal{\begin{align}#1\end{align}}

\def\be#1\ee{\begin{equation}#1\end{equation}}
\def\bea#1\eea{\begin{align}#1\end{align}}
\newcommand{\bear}{\begin{eqnarray}}
\newcommand{\eear}{\end{eqnarray}}

\begin{document}

\title{\bf\large
\vspace*{-2cm}
\begin{flushright}
{\normalsize\normalfont}
\vspace*{1cm}
\end{flushright}
A note on the Loop Formula for the fermionic determinant
}

\author{
Erhard Seiler$^b$\footnote{email: ehs@mpp.mpg.de} and
Ion-Olimpiu Stamatescu$^c$\footnote{email:  
I.O.Stamatescu@thphys.uni-heidelberg.de}  
\mbox{} \\
\mbox{} \\   
 $^b${\em\normalsize Max-Planck-Institut f\"ur Physik 
(Werner-Heisenberg-Institut)} \\
 {\em\normalsize M{\"u}nchen, Germany} \\
$^c${\em\normalsize Institut f\"ur Theoretische Physik, Universit\"at 
Heidelberg and FEST} \\
 {\em\normalsize Heidelberg, Germany}
}

\date{\today}

\maketitle

\abstract{\noindent
A formula expressing the fermionic determinant as an infinite product of 
smaller determinants is derived and discussed. These smaller determinants 
are of a fixed size, independent of the size of the lattice and are 
indexed by loops of increasing length.}

\newpage

\tableofcontents

\newpage

\section{Introduction}

The study of the effects of virtual particles has a very long history. In 
particular the vacuum polarization due to electron-positron pairs was 
studied first by Euler and Heisenberg \cite{Heisenberg:1935qt} and later 
by Schwinger \cite{Schwinger:1951nm}. 

Let us mention some of the later developments: the vacuum polarization to 
all orders is given by the fermion determinant. Its mathematical 
properties were studied later for instance by Seiler \cite{seiler} in the 
context of constructive quantum field theory, where a stability problem 
arising in four dimensional quantum electrodynamics was pointed out. This 
latter issue was studied more extensively by \cite{Haba:1981rp}. 
The study of this issue culminated in the work of Fry \cite{Fry:2015qua} 
which also includes the effect of fermions on non-abelian gauge fields.

The quantum effects on vortex fields were analyzed by Langfeld et al 
\cite{Langfeld:2002vy}, using Schwinger's propert time formalism. Schmidt 
and Stamatescu \cite{Schmidt:2003bf}, also using this formalism, pointed 
out that the fermion (and boson) determinant on the lattice can be viewed 
as a gas of closed loops which can be simulated numerically via a random 
walk.
 
In this note we also consider lattice gauge theories; we derive and 
discuss a general loop formula for the fermion determinant. This formula 
provides a systematic approximation for the fermionic determinants which 
can be used in full QCD analyses. In particular it proved useful in 
problems of QCD at non-zero chemical potential where it allowed to set up 
the so called HD-QCD (for high density quantum chromodynamics) 
approximation for large mass and chemical potential 
\cite{Bender:1992gn,Blum:1996xt, Engels:1999tz} which was used for 
physically relevant simulations in leading (LO) and next-to-leading orders 
(NLO) \cite{Aarts:2001dz,dfss,Fromm:2012uj,Aarts:2015yba}. This formula 
involves, however, some subtleties both in its derivation and in its 
interpretation and we think it is of some interest to address them here.

Let us repeat that the formula itself is not new, but in this note we 
provide some necessary clarifictions of its status.
 
This loop formula is based on the loop expansion derived long ago by one 
of us \cite{Stamatescu:1980br} for the logarithm of the determinant with 
Wilson fermions in an external gauge field. This logarithm is expressed 
there as a power series expansion in the hopping parameter $\kappa$, in 
which each term is given as a sum of closed loops. The loop formula 
proceeds from this expansion and expresses the full determinant as a 
product over determinants of a much smaller fixed dimension (the tensor 
product of spin and color spaces) involving closed loops on the lattice of 
increasing complexity \cite{dfss}. It involves the resummation of nested 
infinite series and therefore its correctness beyond formal algebra 
depends on (absolute) convergence. We will discuss in more detail the 
derivation of the formula as well as its limitations and possible 
misunderstandings in using it, since one might think that the zeroes of 
the determinant are given by the zeroes of the factors of the product. In 
fact this is not the case, as we will explain in the following, and the 
formula should instead be understood as a systematic approximation, as 
illustrated below.

In particular for a finite lattice this formula expresses the 
determinant which is a polynomial of finite order in the hopping 
parameter as an infinite product. Obviously this can make sense only 
where the infinite product converges, which is equivalent to convergence 
of the expansion of its logarithm; this convergence will break down at 
the latest at the first zero encountered, either on the left hand side 
(LHS) or in one of the factors on the right hand side (RHS).

A simple example might help at this point to illustrate both derivation 
and problems of the loop formula. Consider 
\be 
1-\kappa(x+y) = \e^{ \ln (1-\kappa(x+y))}\,.
\ee 
We expand the logarithm in powers of $\kappa$ and $x,y$, but for 
bookkeeping purposes we treat $x,y$ as noncommuting symbols, so we 
consider the string $xxyy$ as different from $xyxy$. This way we get
\beal 
&\ln (1-\kappa(x+y))= -\kappa x-\kappa 
y-\frac{\kappa^2}{2} \left(xx +yy + 2x y) \right)\notag\\ &- 
\frac{\kappa^3}{3}\left(xxx+xxy+ xyx+xyy +yxx+ yxy+ yyx+yyy\right) 
\notag\\ & -\frac{\kappa^4}{4}\left(xxxx + xxxy+\ldots\right)+\ldots\,.  
\label{e.xy1} 
\eeal 
Symbolically we can write 

\be 
\ln (1-\kappa(x+y))=-\sum_l \frac{\kappa^l}{l} \sum_{s_l}s_l (x,y)\,, 
\ee 
where $s_l(x,y)$ stands for strings of lengh $l$ formed from $x$ and 
$y$. Next we introduce `primary strings' $s^P_l$ as those strings that are 
not repetitions of other strings and resum now first over all repetitions 
of primary strings (``s-resummation'' in the following) and then over the 
primary strings. This way we get 
\be 
\ln (1-\kappa(x+y))=-\sum_l 
\sum_{s^P_l}\sum_s \frac{\kappa^{l s}}{l s} s^P_l(x,y)^s\,. 
\ee 
Up to cyclic permutations the primary strings of length 1 are $x$ and 
$y$, of length 2 there is only $xy$, of length 3 there are 
$xxy,yyx,xyy,yxx,xyx,yxy$, of length 4 $xxxy, xxyy$ etc.. Notice that 
$xyxy$ is not a primary string but a repetition of the string $xy$.
We now introduce equivalence classes $[s^P_l]$ for the $l$ strings of 
length $l$ differing by cyclic permutations and obtain 
\beal 
\ln 
(1-\kappa(x+y))&=-\sum_l \sum_{[s^P_l]}\sum_s \frac{\kappa^{l s}}{ s} 
s^P_l(x,y)^s\notag\\ & =\sum_l\sum_{[s^P_l]}\ln (1-\kappa^l s^P_l(x,y))\,. 
\label{reorder} 
\eeal 
 So we 
obtain
\beal 
&\ln (1-\kappa(x+y))=\notag\\ 
&\ln (1 - \kappa x) +\ln (1- \kappa y)  
+\ln (1 - \kappa^2 x y)  +\ln (1 - \kappa^3 x^2 y) +\ln (1 - \kappa^3 x 
y^2) \notag \\ 
&+\ln (1 - \kappa^4 x^3 y)  +\ln (1 - \kappa^4 x y^3) +\ln (1 - \kappa 
^4 x^2 y^2) +\,\ldots\,. \label{e.xy} 
\eeal 
After exponentiation this gives 
\beal 
&1-\kappa(x+y) \notag \\ &= (1 - \kappa x) (1 - \kappa y) (1 - \kappa^2 x y) 
(1 - \kappa^3 x^2 y) (1 - \kappa^3 x y^2)\notag \\
& \times (1 - \kappa^4 x^2 y^2) (1 - \kappa^4 x^3 y) (1 - \kappa^4 x 
y^3)\times \ldots\,. 
\label{e.xy3} 
\eeal 
Regarding now the symbols $x,y$ again as complex numbers we turn to the 
question of convergence. One can see in this example that truncating the 
product by keeping only the factors containing up to $\kappa^4$ the LHS is 
only reproduced up to order $\kappa^5$:
\beal 
&(1 - \kappa x) (1 - \kappa y) (1 - \kappa^2 x y) (1 -\kappa^3 x^2 y) 
(1 -\kappa^3 x y^2) \notag \\ &\times (1 - \kappa^4 x^2 y^2) 
(1 - \kappa^4 x^3 y) (1 - \kappa^4 x y^3) \notag\\ 
&= 1-\kappa(x+y) + \kappa^5 (x^4y + 2x^3y^2 +2x^2y^3 +xy^4) + 
O(\kappa^6)\,.  
\eeal 
Of course this mechanism will operate the same way at any order:  
truncation at order $\kappa^n$ will produce an approximation to the 
determinant up to order $O(\kappa^{n+1})$.

To determine when this expansion converges, note that for $x,y\ge0$ all 
terms in Eq.(\ref{e.xy1}) have the same sign, so this expansion converges 
absolutely for $\kappa<1/(x+y)$.  For general $x,y$ we thus have absolute 
convergence of any reordering and resumming of the expansion for 
$\ln(1-\kappa(x+y))$, provided 
\be 
\kappa<\frac{1}{|x|+|y|}\,, 
\ee 
so in particular the expansion Eq.(\ref{reorder}) converges. 
Exponentiating we obbtain convergence of the infinite product 
Eq.(\ref{e.xy3}).

One can see directly that in this region neither the LHS nor any of the 
factors on the RHS vanishes, so in the region of 
convergence there is no paradox of one side vanishing with the other 
side nonvanishing. The infinite product on the RHS converges in this 
region to the LHS, which provides the analytic continuation to all of 
$\C^2$. 

It should not be surprising that the location of the zeroes 
of the LHS is not well approximated by the zeroes of the factors on the 
RHS: the lowest zeroes as given by the two linear factors, $(1-\kappa 
x)(1-\kappa y)$, e.g., are at $\kappa=1/x$ and $\kappa=1/y$, whereas the 
true zero is at $\kappa=1/(x+y)$. Only after expanding the truncated 
product we obtain a first order approximation of the LHS: $(1-\kappa 
x)(1-\kappa y)= 1 - \kappa(x+y) +O(\kappa^2)$.

In the following we shall first present the principle leading to this 
formula in Sect. \ref{s.nut}, then briefly describe its derivation for QCD 
with Wilson fermions, which involves some subtleties, in Sect. 
\ref{s.der}. Details and illustration of both the derivation of the loop 
formula and of its problems are given in the Appendices.

\section{The idea in a nutshell} 
\label{s.nut}

The principle behind the formula is easy to explain: let
\be
W=\1-\kappa M\,;\quad M_{ii}=0,\; i=1,\ldots n
\ee
be a $N\times N$ matrix  . Then 
\be
\ln \det W= - \sum_{l=1}^\infty \frac{\kappa^l}{l}\tr \,M^l\, 
\label{logexp}
\ee
which converges as long as $||M||<1/\kappa$ or even as long as the 
spectral radius $\rho(M)<1$. We can interprete $\tr M^l$, which 
involves a $l$-fold sum over the matrix indices as a sum over all 
closed paths (journeys or loops) of length $l$ over the the index set. 
To be precise, we define a path of length $l$ as a map
\be
{\cC}_l:\{0,1,\ldots, l\} \rightarrow \{1,2,\ldots N\} 
\ee
with $\cC_l(l)=\cC_l(0)$. Furthermore we define a weight for each 
path
\be
M_{{\cC}_l}= \prod_{i=0}^{l-1} M_{{\cC}_l(i),{\cC}_l(i+1)}\,,
\label{weight}
\ee
so that
\be
\tr\, M^l=\sum_{{\cC}_l}M_{{\cC}_l}\,.  
\ee 
and
\be
\ln \det W= - \sum_{l=1}^\infty \sum_{{\cC}_l}
\frac{\kappa^l}{l}M_{{\cC}_l}\,.
\label{loop0}
\ee
The weight defined in Eq.(\ref{weight}) is clearly invariant under cyclic 
permutations of $\{0,1,\ldots, l\}$, so it makes sense to introduce 
equivalence classes $[\cC_l]$ of paths $\cC_l$  that can be mapped 
onto each other by a cyclic permutation. So we finally can write
\be
\ln \det W= - \sum_{l=1}^\infty \sum_{[{\cC}_l]}\kappa^l 
M_{{\cC}_l}\,.
\label{loop1}
\ee 

As long as the series (\ref{loop0}) converges absolutely, we can make any 
kind of rearrangement, summing first over (finite or infinite) subsets of 
paths; the result will not be affected. For instance we can define 
{\it primary paths} ${\cC}^P_l$ of length $l$ as paths which are 
not repetitions of other paths, and rewrite the formula as
\bear
&&\ln \det W = -\sum_{l=1}^\infty \sum_{{{\cC}^P_l}}
\sum_{s=1}^\infty \frac{\kappa^{sl}}{sl}(M_{{{\cC}^P_l}})^s  \\
&&=- \sum_{l=1}^\infty \sum_{{[{\cC}^P_l]}} 
\sum_{s=1}^\infty \frac{\kappa^{sl}}{s}(M_{{{\cC}^P_l}})^s 
= -\sum_{l=1}^\infty \sum_{[{\cC}^P_l]} \ln(1-\kappa^l 
M_{{\cC}^P_l})\,
\label{loop2}
\eear  
After exponentiating, this gives 
\be
\det W = \prod_{l=1}^\infty \prod_{[{\cC}^P_l]} (1-\kappa^l 
M_{{\cC}^P_l})\,.
\label{prod}
\ee
This last expression looks like a factorization of $\det W$, but it is 
not, because the convergence requirement limits its validity. Absolute 
convergence can hold at most for $|\kappa|<|\kappa_c|$ where $\kappa_c$ is 
the zero of $\det W$ closest to the origin. But of course it may break 
down earlier, because convergence of the logarithmic expansion 
Eq.(\ref{logexp}) does not imply absolute convergence of the path sum 
Eq.(\ref{loop1}): absolute convergence of Eq.(\ref{loop1}) requires
\be
\lim\sup_{l} \left[\sum_{[{\cC}_l]} 
|M_{{\cC}_l}|\right]^{1/l}<\frac{1}{\kappa}
\label{loopconv}
\ee
whereas absolute convergence of Eq.(\ref{logexp}) requires only

\be
\lim\sup_{l} |\tr M^l|^{1/l}<\frac{1}{\kappa}\,,  
\label{logexpconv}
\ee
which is a much weaker condition. In fact the latter can easily be 
estimated in a way independent of the `volume' $N$:
\be
|\tr\, M^l|^{1/l}\le N^{1/l}\rho(M)^{1/l}\le N ||M||\,,
\ee
but it is in general not true that		

\be
\sum_{{\cC}_l}|M_{{\cC}_l}|\le N ||M||\, 
\ee
and therefore it is difficult to find estimates for $|\tr 
M^l|^{1/l}|$ independent of $N$.

So in general there is no simple relation between the zeroes of the 
factors of Eq. (\ref{prod}) and those of $\det W$.

In the next section we turn to the main problem of deriving a similar loop 
formula for the Dirac-Wilson determinant. There the index set is composed
of lattice indices and internal ones (spinor, color, flavor) and the 
definitions of loops etc. are modified accordingly. On the other hand, the 
Dirac-Wilson operator connects only nearest neighbor lattice points, which 
leads to a tremendous simplification, since instead of the `flight 
journeys' considered  here, we have only paths consisting 
of nearest neighbour steps.  

\section{The loop formula for the Dirac-Wilson determinant}
\label{s.der}

We use here the Wilson fermion formulation in $d=2$ or $4$ dimensions 
\bear
W &=& \one- \kappa Q = \one -  \kappa\, \sum_{i=1}^{d-1} \left( 
\Gamma_{+i}\,U_i\,T_i +
\Gamma_{-i}\,T^{-1}_i\,U^{-1}_i\right) \nn \\
&& \ \ \  - \kappa\, \left( \e^{\mu}\,\Gamma_{+4}\,U_4\,T_4 +
\e^{-\mu}\,\Gamma_{-4}\,T^{-1}_4\,U^{-1}_4 \right) ] \, \label{e.W}  \eear
where  $\kappa$ is the hopping parameter,
$T_{\mu}$ are lattice translation, $U_{\mu}$ are link matrices
(assumed here in the fundamental representation of $SU(3)$ or $SL(3,C)$), and
\bear
\Gamma_{\pm \mu} &=& \one \pm \gamma_{\mu},\ \ \gamma_{\mu}=\gamma_{\mu}^*,\ \
\gamma_{\mu}^2=\one  \label{e.G1} \\  
\tr \Gamma_{\pm \mu} &=&d, \ \ \det\Gamma_{\pm \mu}=0 , \  \
 \Gamma_{\pm \mu}\Gamma_{\mp \mu} =0, \ \ \Gamma_{\pm \mu}^2 = 
 2\Gamma_{\pm \mu} \label{e.G2}
\eear
are  Wilson's projectors for  the Dirac d.o.f.'s. In this formulation the 
loop formula takes a simpler form, since back-steps are forbidden. 
See also Appendix A.

The loop formula \cite{Stamatescu:1980br,Aarts:2001dz,dfss} for a 
finite simple square ($d=2$) or hypercubic ($d=4$) lattice reads:
\bear
\det W &=& \det( \one- \kappa Q) = {\rm exp} ( \tr 
\ln (\one - \kappa Q) )     \label{e.hopg0} \\
 &=&  {\rm exp} \left[-\sum_{l=1}^\infty  \sum_{\left\{{\cal
C}_l\right\}} \sum_{s=1}^\infty ~{{{ g_{{\cal C}_l}}^s}\over
s}\,\Tr  \left[{\cal L}_{{\cal C}_l}^s\right] \right]  \label{e.hopg1}\\
 &=& 
\prod_{l=1}^{\infty} ~\prod_{\left\{{\cal C}_l\right\}}  
  \det_{\rm D,C} \left(\one~-~ g_{{\cal C}_l}
{\cal L}_{{\cal C}_l}\right) \, \label{e.hopg2}\,, \\
g_{{\cal C}_l} &=& \kappa^l\,\left(\epsilon \, \e^{ N_{\tau}\mu}\right)^r\, . 
\eear
Here  $r$ is the net winding number of the path in the time direction,
with periodic or antiperiodic b.c. and $\epsilon = +1(-1)$ correspondingly.
We assume periodic b.c. in the `spatial' directions. 
Notice that the terms in $Q$ imply unit steps on the lattice, therefore 
a lattice path ${\cal C}_l$ in Eq. (\ref{e.hopg2}) is a closed chain of 
lattice points as produced by the $l$th power of $Q$ in Eq. 
(\ref{e.hopg0}).
Due to the trace condition 
a path must close, but it can 
 repeatedly visit the same lattice point
 before closing (the Pauli principle was used in writing the determinant, after
that everything is only matrix algebra). 
Notice that this condition also implies that on an 
 even lattice
 $l$ is even. In the following we shall speak of the  ${\cal C}_l$ paths in 
 Eqs. (\ref{e.hopg1}-\ref{e.hopg2}) as
{\it primary paths}. They are 
 {\it distinguishable, 
non-exactly-self-repeating lattice 
closed paths of length $l$}.  
Since the primary paths are closed paths and the ensuing $\Tr$
do not depend on cyclic permutations of their points
we do not identify a starting point for ${\cal C}_l$. Nonetheless, in producing 
such a path we can start at each of its points, therefore
 we get a multiplicity factor $l$.
 
  We define the loops ${\cal L}_{{\cal C}_l}$ as 
the  chains of links
and $\Gamma$ factors on the primary path ${\cal C}_l$ and call them
{\it primary loops}. A loop, however, may be repeated 
in covering the  path ${\cal C}_l$ before the Dirac and colour
traces of $\Tr$ close, and the exponent 
$s$ in Eq. (\ref{e.hopg1}) counts the repetitions in covering  ${\cal C}_l$. 
Since stepping further after completing the primary path we only 
obtain 
identical paths, however, these repetitions do not 
produce a further multiplicity
factor, therefore no further factor $s$ counting the
repetitions appears. 
These considerations have 
  been taken into account
in Eqs. (\ref{e.hopg1},\ref{e.hopg2}).

 With $\lambda$
 denoting the links along  ${\cal C}_l$ we have;
\bear
{\cal L}_{{\cal C}_l} &=& \prod_{\lambda \in {\cal C}_l}\Gamma_{\lambda}
U_{\lambda} = \Gamma_{{\cal C}_l}\, U_{{\cal C}_l}, \quad 
\Gamma_{{\cal C}_l}=\prod_{\lambda \in {\cal C}_l}\Gamma_{\lambda},\quad 
U_{{\cal C}_l} = \prod_{\lambda \in {\cal C}_l} U_{\lambda}\,. \label{e.lp1} 
\label{e.lp2}\eear
 The
Dirac factors $\tr_D \Gamma_{{\cal C}_l}$ 
can be calculated for each ${\cal C}_l$ geometrically \cite{Stamatescu:1980br} 
or numerically. 

The contribution of 1- and 2-dimensional loops ${\cal L}_{{\cal \tilde C}_l}\, ,\ \
{\cal \tilde C}_l\,:$ linear and planar loops closing on or over the 
lattice, in
(\ref{e.hopg2})
 further simplifies to
\bear
&&\prod_{l=1}^{\infty} ~\prod_{\left\{{\cal \tilde C}_l\right\}}  
  \det_{\rm C} \left(\one~-~ h_{{\cal \tilde C}_l}
{\cal L}_{{\cal \tilde C}_l}\right)^{d/2} \, \label{e.hopg3},\\
&=&\prod_{l=1}^{\infty} ~\prod_{\left\{{\cal \tilde C}_l\right\}}  
  \left(1~-~ h_{{\cal \tilde C}_l}\tr_C U_{{\cal \tilde C}_l} 
  ~+~ h_{{\cal \tilde C}_l}^2\tr_C U_{{\cal \tilde C}_l}^{-1}~-~ 
  h_{{\cal \tilde C}_l}^3
\right)^{d/2} \, \label{e.hopg4}\\
h_{{\cal C}_l} &=& g_{{\cal C}_l}\times
\frac{2}{d}\tr\Gamma_{{\cal C}_l} = \kappa^l\,\left(\epsilon \, 
\e^{N_{\tau}\mu}\right)^r\times
\frac{2}{d}\tr\Gamma_{{\cal C}_l}, \ 
\det \Gamma_{{\cal C}_l} =0 \label{e.lp3}
\eear
where Eq. (\ref{e.hopg4}) holds for $SL(3,C)$ ($SU(3)$) gauge group.

In deriving Eqs. (\ref{e.hopg3},\ref{e.hopg4}) we used the fact
that the Dirac and Yang-Mills traces in Eq. (\ref{e.hopg1}) factorize and 
the former concern products of the projection operators Eqs. (\ref{e.G1},
\ref{e.G2}):
\bear
&&\Tr  \left[{\cal L}_{{\cal C}_l}^s \right] = 
\tr \left[ \Gamma_{{\cal C}_l}^s\right]
 \tr \left[ U_{{\cal C}_l}^s\right]
, \label{e.lp4} \\ 
&&\frac{2}{d}\,\tr\left[\Gamma_{{\cal C}_l}^s \right] = 
\left[\frac{2}{d}\,\tr \Gamma_{{\cal C}_l} \right]^s\,, 
\label{e.lp5}
\eear
where the second equation holds for linear and planar loops
 \cite{Stamatescu:1980br}.
In particular, e.g. for straight Polyakov loops and their inverses
${\cal L}_{{\cal C}_{N_{\tau}}} = 
{ P}$,  ${\cal L}^{-1}_{{\cal C}_{N_{\tau}}} = 
{ P}^{-1}$  we have
\bear
h_{ P}   
= (2\zeta)^{N_{\tau}},\  h_{{ P}^{-1}}  =  
 (2{\tilde \zeta})^{N_{\tau}},\ \ 
 \zeta= \kappa \e^{\mu},\ {\tilde \zeta}=
\kappa \e^{-\mu}.\label{e.zeta}
\eear
For loops which explore more than two dimensions of the lattice Eq. 
(\ref{e.lp5}) does not hold generally and therefore Eq. (\ref{e.hopg2})
cannot always be rewritten similar to Eq. (\ref{e.hopg4}).
 
We shall speak of {\it basic paths} as 
primary paths without repeated visitation of a point. 
Notice that the number of basic paths is finite on 
a finite lattice. The primary paths are obtained by chaining basic paths 
in arbitrary order and arbitrarily often repeated 
(only the resulting chain as a whole should
not be repeated). Hence even on a finite lattice 
there are infinitely many primary paths.
 
Notice therefore that the RHS of the loop formula on any lattice 
is an infinite product.  
Since on a finite lattice the determinant is a polynomial in $\kappa$ of
 order $N_{max} = d N_L N_c$ with $d=2,4$ the dimension, 
$N_L$ the lattice volume, $N_c$ the number of colors the loop formula implies 
cancellations of the higher orders. This happens algebraically 
(i.~e. without worrying about convergence), but it is justified 
analytically only if the nested infinite series involved converge 
absolutely. See the discussion in sect. 2.

On the other hand 
 arbitrarily truncating the RHS by keeping only a finite 
number of factors up to some $l_0$ will make the loop formula into an
approximation of order $l_0 $. The zeroes suggested by the RHS 
are therefore not true zeroes of the determinant but will corroborate
to provide the approximation of the latter 
 to the above order.

When we are dealing with finite temperature and $\mu > 0$, with
(anti-)periodic 
boundary conditions in time the coefficients for loops with
 positive net winding number $r>0$ in the time direction will 
 have coefficients containing powers of $\zeta$ 
 and can therefore be
 of order 1 or larger: 
\bear
g_{{\cal C}_l} = \kappa^l \left(e^{N_{\tau}\mu}\right)^r = 
\kappa^{l_{\sigma}} \zeta^{r\,N_{\tau}},\,\ l_{\sigma}
= l-r\,N_{\tau} \geq 0 ,\, \ \zeta=\kappa \e^{\mu} \label{e.zeta2}
\eear
It is useful therefore in this case 
to reorder the loops  
according to powers $\kappa^{l_{\sigma}}$, the 
lowest order (LO) being given by 
straight Polyakov loops, the higher orders will involve Polyakov 
loops `decorated' by spatial excursions.  To LO and NLO we can use 
Eqs. (\ref{e.hopg4},\ref{e.lp3},\ref{e.zeta}) for these loops, see 
\cite{Aarts:2001dz,dfss} for details. Notice that it is important to
use here the s-resummation which brings the loops into
the determinant factors, in order to avoid large numbers in the exponents. 
Loops without windings have no $\zeta$ factors and start at order $\kappa^4$.

For completeness the 
formal derivation is described in detail in Appendix \ref{app1}.
In Appendix \ref{app2} we illustrate on two simple examples the evaluation
of the formula and the problem of the zeroes.

\section{Conclusions}

The loop formula  Eqs. (\ref{e.hopg0}-\ref{e.hopg2})
 expresses the determinant of the ${\cal N}\times{\cal N}$ Wilson 
fermionic matrix, with ${\cal N}=N_{Lattice}\times N_{Dirac}\times 
N_{Colour}$, as an infinite product of determinants of a fixed 
(independent of $N_{Lattice}$), small dimension $\left(N_{Dirac}\times 
N_{Colour}\right)^2$. The loop formula is based on the loop expansion 
given in \cite{Stamatescu:1980br}, which itself represents a hopping 
parameter expansion, and its derivation requires careful reordering of the 
terms and resummation of infinite nested series. This way one obtains an 
algebraically correct representation, whose use as an approximation 
requires, however, a convergence analysis.
 
The infinite product of small determinants on the RHS can be ordered 
according to the length of the loops or to the order of the hopping 
parameter. The individual zeroes of the RHS are not zeroes of the 
LHS but the RHS converges to the big, exact determinant of the LHS in 
the region below the lowest zeroes.  Truncation of the RHS after a given 
loop length produces approximations of the LHS after expanding the product 
and retaining the terms up to the corresponding order.

The formula provides a valid series of approximations in its range of 
absolute convergence; this domain does, however, not include the zeroes of 
the full determinant nor those of any approximants. It can thus not be 
used for an approximate determination of the zeroes of the full 
determinant, but the approximants can still be useful, e.g. in problems of QCD 
at finite chemical potential (or density). The formula may be interpreted 
as providing an ensemble of loops; this suggests that higher orders, which 
are hard to evaluate algebraically, might be instead be produced by 
stochastic generation of loops as in \cite{Schmidt:2003bf}.

\newpage
\appendix
\section{Grand Canonical Partition function for QCD 
with Wilson fermions at $\mu>0$}
\bea
&{\cal Z}(\beta,\kappa,\gamma_G,\gamma_F,\mu) 
= \int[DU]\, 
\e^{-S_G(\beta,\gamma_G,\{U\})}{\cal Z}_F({ {\kappa}},\gamma_F, \mu, \{U\}) \, ,
\label{e.gcpt} \\ 
&S_G(\beta,\gamma_G,\{U\})   
= -\frac{\beta}{N_c}\,Re\,\tr\,\left(\frac{1}{\gamma_G}\,
\sum_{j>i=1}^3\,
P_{ij} + \gamma_G\,\sum_{i}\,P_{i4}\right)\, , \label{e.YMa}\\ 
&{\cal Z}_F({ {\kappa}}, \gamma_F,\mu, \{U\}) =  
\det W ({ {\kappa}}, \gamma_F,\mu, \{U\}) \, , \label{e.det}\\  
&W = \one- \kappa Q = \one -  \kappa\, \sum_{i=1}^3 \left( 
\Gamma_{+i}\,U_i\,T_i +
\Gamma_{-i}\,T^{-1}_i\,U^{-1}_i\right) \nn \\
& \ \ \  - \kappa\, \gamma_F\, \left( \e^{\mu}\,\Gamma_{+4}\,U_4\,T_4 +
\e^{-\mu}\,\Gamma_{-4}\,T^{-1}_4\,U^{-1}_4 \right)  \, \label{e.WW}  \\
&\Gamma_{\pm \mu} = \one \pm \gamma_{\mu},\ \ \gamma_{\mu}=\gamma_{\mu}^*,\
\gamma_{\mu}^2=\one ,\, \tr \Gamma_{\pm \mu} =4 ,\, \det\Gamma_{\pm \mu}=0 \\ 
&\kappa =
\frac{1}{2(M+3+\gamma_F\,\cosh \mu)} = \frac{1}{2(M_0+3+\gamma_F)} \, ,
\eea
\no (one flavour). Here $M$ is the ``bare mass'', $M_0$  the bare mass at $\mu=0$,
 $U$  denote the link variables and $T$ 
lattice translations. For completeness we introduced anisotropy factors 
$\gamma_G,\gamma_F$ which have to be
 tuned, by requiring isotropy of physical quantities at $T=\mu=0$
 (hadron masses, string tension etc), defining 
 so a unique physical anysotropy of the lattice spacings
 $\eta = a_{\sigma} / a_{\tau}$ which will enter the physical temperature, 
 chemical potential, masses, etc. In the following $\gamma_G=\gamma_F=1$.
The exponential prescription for $\mu$ ensures cancelling of 
divergences in the small $a$ limit.

\section{Formal derivation}
\label{app1}

\begin{figure}
\begin{center}
\epsfig{file=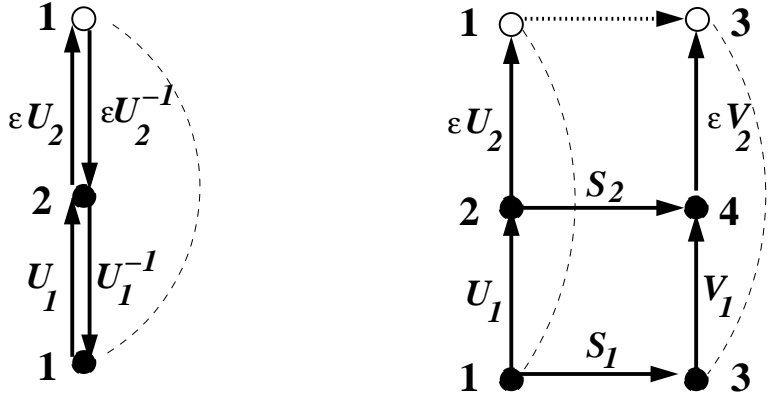, width=10cm}
 \caption{ Case 1 and case 2, $\epsilon =1(-1)$ for periodic boundary 
conditions (pbc) (apbc), repectively. On the right plot the inverse links are 
no longer indicated. 
 }
\label{f.loop22}
\end{center}
\end{figure}

In the following we shall formally derive the loop formula. For illustration 
we shall refer to Fig. \ref{f.loop22}. With 
\be
W = \one- \kappa\, Q
\ee
we have at first step
\bear
\tr \ln W  &=& \tr \ln ( \one- \kappa Q) = -\sum_{l=1}^\infty \frac{\kappa^n}{n}
 \tr
 \left(Q^n \right) \label{e.hopg5}\\
 &=& -\sum_{l=1}^\infty  \sum_{\left\{{\cal
C}_l\right\}} \sum_{s=1}^\infty ~{{{  g_{{\cal C}_l}}^s}\over
s}\,\Tr  \left[ {\cal L}_{{\cal C}_l}^s\right]  \label{e.hopg6}
\eear

Notice: 

(1) Due to the traces we obtain only connected, closed loops 
on the lattice (without backtracking for Wilson fermions).

(2) We consider now {\it primary loops}  $ {\cal L}_{{\cal C}_l}$.
 They are of all possible lengths $l$ and may close 
on or over the toroidal lattice. 

{\it Example}: In $\frac{1}{4}\tr Q^4$ the loop $\Tr S_1V_1S_2^{-1}U_2$ 
appears with multiplicity
4 (since it can be started at each node) while leading each time 
to the same contribution by cyclicity of $\Tr$
- cf Fig.(\ref{f.loop22}) right.

(3) Each primary loop can appear repeated any number of times $s$. As already pointed out, 
because of
the  indistiguishability in the repetition there is no further multiplicity.

 {\it Example}: 
On a 1-dim periodic lattice of length 2  the term $\frac{1}{2}\tr Q^2$ has 2 
  loops  closing over the lattice, $U_1U_2$ and $U_2U_1$, contributing equally
 $\Tr U_1U_2 = \Tr U_2U_1 $, see
Fig. \ref{f.loop22} left. Since the corresponding paths are identical 
we speak of them as one primary loop with 
multiplicity 2, contributing thus $2 \times \frac{1}{2} \Tr U_1U_2$
in the loop formula. 

The term $\frac{1}{4}\tr Q^4$ of order $n=4$ contains
$(U_1+U_2)^4 $ and has just 2 different
 terms, $U_1 U_2 U_1U_2 $ and $ U_2 U_1 U_2 U_1$, 
 allowed by the lattice trace (notice
 that a term like $U_1^2 U_2^2$ cannot 
 appear since each link is associated with 
 a lattice step).  One can see $ U_2 U_1 U_2 U_1$ as the first 
 permutation of $ U_1 U_2 U_1U_2  $, 
 further permutations do not bring new loops. 
 The contributions of these two loops to the ensuing $\Tr$ 
 are equal 
 by cyclicity of the trace. Therefore we speak here of just one loop, say 
 $U_1 U_2 U_1U_2$.
 Since it is a
 genuine repetition it does not represent a primary loop, 
 but the
 $s=2$ repetition $(U_1U_2)^2$ of the primary loop $U_1U_2$ of
  multiplicity 2 and contributes therefore 
  $2 \times \frac{1}{4}\Tr \left[(U_1U_2)^2\right]$ in the loop sum.

(4) Identification of primary loops. 

{\it Example}: The loops $S_1V_1S_2^{-1}U_1^{-1}$ and
$S_1V_1S_2^{-1}U_2$ are basic loops, the first one is a plaquette, the second
a decorated Polyakov loop, both appearing with 
multiplicity 4.
$S_1V_1S_2^{-1}U_1^{-1}S_1V_1S_2^{-1}U_1^{-1}S_1V_1S_2^{-1}U_2$
 is a primary loop, consisting of
the iterated basic loop Plaquette and the basic Polyakov loop and 
has multiplicity 12. In contrast to this, the loop 
$S_1V_1S_2^{-1}U_1^{-1}S_1V_1S_2^{-1}U_1^{-1}S_1V_1S_2^{-1}U_1^{-1}$ 
is not a primary loop but 
the $s=3$ repetion of the basic loop plaquette $S_1V_1S_2^{-1}U_1^{-1}$ 
and appears 
therefore with multiplicity 4.
\medskip

We now make  the 2-nd step to obtain:

\bea 
\det W =& {\rm exp} ( \tr \ln W )  \label{e.hopg7}\\
 =& 
\prod_{l=1}^{\infty} ~\prod_{\left\{{\cal C}_l\right\}} ~ 
  \det_{\rm D,C} \left(\one~-~ g_{{\cal C}_l}
{\cal L}_{{\cal C}_l}\right) \, \label{e.hopg8}
\eea

Notice:

(5) The summands in Eq. (\ref{e.hopg6}) are Dirac and colour traces, that is
just complex numbers.
 Each summand corresponding to a primary loop is of the form
\be
 -\sum_{s=1}^\infty ~\frac{  g_{{\cal C}_l}^s}{s}\,\Tr     
\left[{\cal L}_{{\cal C}_l}^s\right] = 
\Tr \ln \left(\one - {{  g_{{\cal C}_l}}}\,
{\cal L}_{{\cal C}_l}\right) \label{e.hopg9}
\ee
since this is just the expansion of a logarithm ({\it s-resummation}). 
It is the repeated covering of primary loops which resumes to the logarithms.

(6)  Using Eq. (\ref{e.hopg9}) we can rewrite
 Eq. (\ref{e.hopg6}) as
\bea 
-\sum_{l=1}^\infty  \sum_{\left\{{\cal
C}_l\right\}} \Tr \ln \left(\one - {{ g_{{\cal C}_l}}}\,
{\cal L}_{{\cal C}_l}\right) \label{e.hopg10}
\eea
Since the summands 
 commute we can exponentiate and
rewrite Eq.(\ref{e.hopg6}) as
 
\bea
\prod_{l=1}^{\infty} ~\prod_{\left\{{\cal C}_l\right\}} \e^{
\Tr \ln \left(\one - {{  g_{{\cal C}_l}}}\,
{\cal L}_{{\cal C}_l}\right)} 
=\prod_{l=1}^{\infty} ~\prod_{\left\{{\cal C}_l\right\}}
\det_{\rm D,C} \left(\one~-~ g_{{\cal C}_l}
{\cal L}_{{\cal C}_l}\right) \,,
\eea
by inverting the general formula Eq.(\ref{e.hopg9}).
This is Eq. (\ref{e.hopg2}).
\bigskip

As remarked before, the zeroes suggested by the RHS factors of the 
loop formula generally are not zeroes of the LHS,
 but keeping the lowest order factors up to some order may provide reasonable
approximations for the LHS, depending on the parameters and configuration.

A particular situation occurs for the case of large mass and chemical 
potential. Here the temporal loops disentangle in the limit $\kappa 
\rightarrow 0,\, \mu \rightarrow \infty,\, \zeta = \kappa \e^{\mu} = 
fixed$ and the determinant reduces to the ``HD-QCD" determinant in LO, 
here for antiperiodic boundary conditions (apbc) ($\epsilon =-1$)

\bear
D_{HD}^{[0]} &=& \prod_{\vec x} \det_C(1+ C P_{\vec x})^{d/2}\det_C(1+ {\tilde C} 
P_{\vec x}^{-1})^{d/2} 
\, \label{e.d0}\\
&=&(1+C\tr P_{\vec x}+ C^2 \tr P_{\vec x}^{-1} +C^3)^{d/2} \nn \\
&&\times  (1+{\tilde C}\tr P_{\vec x}^{-1}+ {\tilde C}^2 \tr P_{\vec x} +
{\tilde C}^3)^{d/2}\\
 g_P &\equiv& C = \left(2 \kappa \e^{\mu}\right)^{N_t},\, g_{P^{-1}} \equiv   
{\tilde C}=\left(2 \kappa \e^{-\mu}\right)^{N_t} 
\eear
Notice that since in  this case the expansion parameter  
$C$ can be of order 1, 
the s-resummation to Eq. (\ref{e.hopg2}) is important
for avoiding large
exponents in Eq. (\ref{e.hopg1}).

\section{Simple examples}
\label{app2}

In the following we treat two examples which both illustrate the loop formula
and its problems.

We use\\
\bea
&\Gamma_{\pm,1} = \one \, \pm \, \sigma_1  
=\left(\begin{array}{cc}1&\pm 1\\\pm 1&1  \end{array}\right) \, ,\quad 
\Gamma_{\pm,2} = \one \, \pm \, \sigma_2 
=\left(\begin{array}{cc}1&\mp i\\\pm i&1  \end{array}\right) \label{e.17}\\ 
&\Gamma_{\pm,i}^2 =2\, \Gamma_{\pm,i}\, , \quad \Gamma_{\pm,i}\Gamma_{\mp,i}=0
\, , \quad \tr\Gamma_{\pm,i}=2\, , \quad \det\Gamma_{\pm,i}=0\\
&\Gamma_{\epsilon_1,1}\, \Gamma_{\epsilon_2,2 } =
- \Gamma_{\epsilon_2,2}\, \Gamma_{\epsilon_1,1 }+2 \Gamma_{\epsilon_1,1}
+2 \Gamma_{\epsilon_2,2 } -2\, \one \\
&= \Gamma_{\epsilon_2,2}\, \Gamma_{-\epsilon_1,1 }+
2\, \Gamma_{-\epsilon_1,1 } -2\, \one\\
&\zu = \kappa \e^\mu \, ,\quad \zd = \kappa \e^{-\mu}\, ,\quad 
\zu \zd = \kappa^2
\eea 
mimicking a 2-dim lattice theory at finite $\mu$.
We use U(1) links to reduce the dimensionality of $W$. 

Notice that we have
\bea
&{\cal L}_{{\cal C}_l} = \Gamma_{{\cal C}_l} 
\prod_{\lambda \in {\cal C}_l}
U_{\lambda}\, , \ \Gamma_{{\cal C}_l} = \prod_{\lambda \in {\cal C}_l}
\Gamma_{\lambda},\, \ \Gamma_{\lambda}=  \Gamma_{\pm,i}\, ,
\  \det \Gamma_{{\cal C}_l} = 0
\\
&\det_{\rm D,C} \left(\one~-~g_{{\cal C}_l}
{\cal L}_{{\cal C}_l}\right)= 1 - h_{{\cal C}_l}
{\cal L}_{{\cal C}_l} ,
\eea
see also Eq. (\ref{e.lp3}).
Since we are interested in chemical potential problems we keep $\zu$ fixed 
in this illustrations and consider various $\kappa$ orders.

\subsection{Case (1)}

\begin{figure}
\begin{center}
\epsfig{file=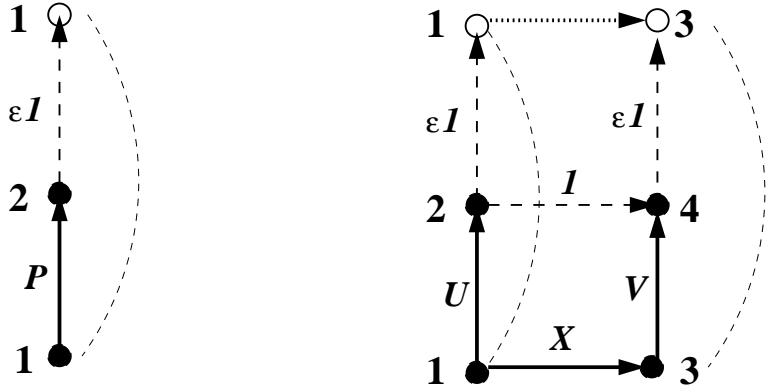, width=10cm}
 \caption{ Case 1 and case 2, maximal gauge fixing, 
 $\epsilon =1(-1)$ for pbc (apbc). Inverse links 
 are not indicated on the plots.}
\label{f.loop22uv}
\end{center}
\end{figure}

We consider a Polyakov loop $P = U_1\,U_2$

\bea
W=
\eea
\[\left(\begin{array}{cccc}
1&	0&	\zu U_1 +\epsilon \zd U_2^{-1}& \zu U_1 - \epsilon \zd U_2^{-1}\\
0&	1&	\zu U_1 - \epsilon \zd U_2^{-1}&	\zu U_1 + \epsilon \zd U_2^{-1}\\
\epsilon \zu U_2 + \zd U_1^{-1}&	\epsilon \zu U_2 - \zd U_1^{-1}&	1&	0\\
\epsilon \zu U_2 - \zd U_1^{-1}&	 \epsilon \zu U_2+ \zd U_1^{-1}&	0&	1
\end{array}\right) \]

Since the determinant is gauge invariant we use gauge transformations 
to put the Polyakov loop on one link, Fig. \ref{f.loop22uv} left plot:
\bea
W=
\eea
\[\left(\begin{array}{cccc}
1&	0&	\zu P +\epsilon \zd 1& \zu P - \epsilon \zd 1\\
0&	1&	\zu P - \epsilon \zd 1&	\zu P + \epsilon \zd  1\\
\epsilon \zu 1 + \zd P^{-1}&	\epsilon \zu 1 - \zd P^{-1}&	1&	0\\
\epsilon \zu 1 - \zd P^{-1}&	 \epsilon \zu 1+ \zd P^{-1}&	0&	1
\end{array}\right) \]

E.g., with $\epsilon =-1$ for apbc simple algebra gives
\bear
\det W &=& 1+ 4\zu^2 P +4\zd^2 P^{-1} +16 \kappa^4 =1+ 4\zu^2 P +O(\kappa^4),
\label{det.case1}\\
 &&P=\tr U_1U_2,
\ P^{-1} 
=\tr U_2^{-1}U_1^{-1} \nn
\eear
It can have real zeroes if $P,\, P^{-1}$ have large, negative real parts.
The primary loops are $P=U_1U_2$ and $P^{-1}=U_2^{-1}U_1^{-1}$ 
and the loop formula Eq.(\ref{e.hopg2}) 
{\it with $s$-resumation} gives 
\bea
\det W =& \e^{ \ln(1+4\zu^2 P) + \ln(1+4\zd^2 P^{-1})}
 = (1+4\zu^2 P)(1+4\zd^2 P^{-1}) \nn \\
 =&1+ 4\zu^2 P +4\zd^2 P^{-1} +16 \kappa^4
\eea
which coincides with the exact result Eq. (\ref{det.case1}).  
Notice that without $s$-resummation the higher orders do not cancel and
we would have obtained wrong results already at 0-th order in $\kappa$: 
\bea
s=1:\ &\e^{ 4\zu^2 P+4\zd^2 P^{-1}}\\
&=
1+ 4\zu^2 P  + 8 \zu^4 P^2 +4\zd^2  P^{-1} +16 \kappa^4 + 8 \zd^4 (P^{-1})^2 
 + \dots \nn \\
s=1+2:\ &\e^{ 4\zu^2 P - 8 \zu^4 P^2 +4\zd^2 P^{-1} 
- 8 \zd^4 (P^{-1})^2}\\
&= 
1+ 4\zu^2 P - \frac{80}{3}\zu^6 P^3+ 4\zd^2  P^{-1} +16 \kappa^4 - 
\frac{80}{3}\,\zd^6 (P^{-1})^3 
+ \dots \nn
\eea

\subsection{Case (2)}

We consider 2 short, connected Polyakov loops (a $2 \times 2$ lattice with pbc
in 1 direction and free b.c in the other). See Fig. \ref{f.loop22}, right plot and Table 1.  
Using again gauge transformations
to a maximal gauge fixing we obtain Fig. \ref{f.loop22uv}, right plot and Table 2. We have  
$\epsilon = 1(-1)$ for pbc(apbc). 
\begin{sidewaystable}
{\small 
\[\left(\begin{array}{cccccccc}
1&0& \zu U_1 + \zd \epsilon U_2^{-1}& \zu U_1 - \zd \epsilon U_2^{-1}&\kappa S_1 &-i\kappa S_1&0&0\\
0&1& \zu U_1 - \zd \epsilon U_2^{-1}& \zu U_1 + \zd \epsilon U_2^{-1}&i\kappa S_1 &\kappa S_1&0&0\\
\zu \epsilon U_2+\zd U_1^{-1} &\zu \epsilon U_2-\zd U_1^{-1}&1&0&0&0& \kappa S_2&-i \kappa S_2\\	
\zu \epsilon U_2-\zd U_1^{-1} &\zu \epsilon U_2+\zd U_1^{-1}&0&1&0&0& i\kappa S_2& \kappa S_2\\	
\kappa S_1^{-1} &i\kappa S_1^{-1}&0&0&1&0& \zu V_1 + \zd \epsilon V_2^{-1}& \zu V_1 - \zd \epsilon V_2^{-1}\\
-i\kappa S_1^{-1} &\kappa S_1^{-1}&0&0&0&1& \zu V_1 - \zd \epsilon V_2^{-1}& \zu V_1 + \zd \epsilon V_2^{-1}\\
0&0& \kappa S_2^{-1}&i\kappa S_2^{-1}&\zu \epsilon V_2+\zd V_1^{-1} &\zu \epsilon V_2-\zd V_1^{-1}&1&0\\
0&0&-i\kappa S_2^{-1}& \kappa S_2^{-1}&\zu \epsilon V_2-\zd V_1^{-1} &\zu \epsilon V_2+\zd V_1^{-1}&0&1\\
\end{array}\right) \]
}
\caption{Fermionic matrix $W$ from Fig. \ref{f.loop22}, right plot. }\bigskip\bigskip
{\small
\[\left(\begin{array}{cccccccc}
0&0& \zu U + \zd \epsilon & \zu U - \zd \epsilon&\kappa X &-i\kappa X &0&0\\
0&0& \zu U - \zd \epsilon & \zu U + \zd \epsilon&i\kappa X &\kappa X &0&0\\
\zu \epsilon +\zd U^{-1} &\zu \epsilon -\zd U^{-1}&0&0&0&0& \kappa &-i \kappa \\	
\zu \epsilon -\zd U^{-1} &\zu \epsilon +\zd U^{-1}&0&0&0&0& i\kappa & \kappa \\	
\kappa X^{-1} &i\kappa X^{-1} &0&0&0&0& \zu V + \zd \epsilon & \zu V - \zd \epsilon \\
-i\kappa X^{-1} &\kappa X^{-1}&0&0&0&0& \zu V - \zd \epsilon & \zu V + \zd \epsilon \\
0&0& \kappa &i\kappa&\zu \epsilon+\zd V^{-1} &\zu \epsilon-\zd V^{-1} &0&0\\
0&0&-i\kappa & \kappa&\zu \epsilon-\zd V^{-1} &\zu \epsilon +\zd V^{-1} &0&0\\
\end{array}\right) \]
}
\caption{Maximal gauge $W\,-\,1 \,=\, - \kappa \, Q$, see 
Fig. \ref{f.loop22uv}, right plot.}
\end{sidewaystable}

\medskip

\noindent Basic and primary loops of length up to $l=6$,
  with corresponding 
 Dirac coefficients $C_i$ and multiplicities $m_i$:
{\scriptsize
\begin{align}
&l=2 :\\
&1-2--1 :  & U_1 U_2 =&L_1\,, \quad &C_1 = 4 \epsilon \zu^2 &\,, \quad m_1=2 \nn\\ 
&3-4--3 : & V_1 V_2 =&L_2 \, , \quad &C_2 = 4 \epsilon \zu^2 &\,, \quad m_2=2\nn\\
&l=4 :\\
&1-3-4-2-1 : &S_1 V_1 S_2^{-1} U_1^{-1} =& L_3\,, \quad &C_3=-4 \kappa^4 &\,, \quad m_3=4\nn\\ 
&2-4--3-1--2 :  &S_2 V_2 S_1^{-1} U_2^{-1}  =& L_4 \,, \quad &C_4 =-4 \kappa^4 &\,, \quad m_4=4\nn\\
&1-3-4-2--1 :  &S_1 V_1 S_2^{-1} U_2 =& L_5 \,, \quad &C_5 = 4 \epsilon  \zu^2\kappa^2 &\,, \quad m_5=4\nn\\
&2-4--3-1-2 :  &S_2 V_2 S_1^{-1} U_1 =& L_6 \,, \quad &C_6 = 4 \epsilon  \zu^2\kappa^2 &\,, \quad m_6=4\nn\\
&l=6 :\\
&1-3-4-2--1-2--1 :  &S_1 V_1 S_2^{-1} U_2 U_1 U_2=& L_7 \,, \quad &C_7 = -16   \zu^4\kappa^2 &\,, \quad m_7=6\nn\\
&2-4--3-4-2--1-2 :  &S_2 V_2 S_1^{-1} U_1 U_2 U_1 =& L_8 \,, \quad &C_8 = -16   \zu^4\kappa^2 &\,, \quad m_8=6\nn\\
&1-3-4--3-4-2--1 :  &S_1 V_1 V_2 V_1 S_2^{-1} U_2=& L_9 \,, \quad &C_9 = -16   \zu^4\kappa^2 &\,, \quad m_9=6\nn\\
&2-4--3-4--3-1-2 :  &S_2 V_2 V_1 V_2 S_1^{-1} U_1  =& L_{10} \,, \quad &C_{10} = -16   \zu^4\kappa^2 &\,, \quad m_{10}=6\nn\\
&1-3-4--3-1-2--1 :  &S_1 V_1 V_2 S_1^{-1} U_1 U_2=& L_{11} \,, \quad &C_{11} = -16   \zu^4\kappa^2 &\,, \quad m_{11}=6\nn\\
&2-4--3-4-2--1-2 :  &S_2 V_2 V_1 S_2^{-1} U_2 U_1 =& L_{12} \,, \quad &C_{12} = -16   \zu^4\kappa^2 &\,, \quad m_{12}=6\nn
\end{align}
}
(and their inverses). 
\footnote{We thank Dr. Elmar Bittner for writing the program
to produce basic and primary loops.}
\medskip

 The loop formula gives:
 \bea
 \det W = \prod_{i=1}^{\infty} (1-C_i L_i)\,(1-C_i L_i^{-1}) \label{e.dloop2}
 \eea
where $L_i$ are primary loops, obtained by chaining together basic loops.
\\

For calculations we can use the maximal gauge,
 setting $U_2=V_2=S_2=1$, see Table 2. Then $L_1=U,\, L_2=V$ are the two 
 straight Polyakov loops,
  $XVU^{-1}$ and $X$ the two plaquettes $L_3, L_4$ and
 $L_5=V\,X,\, L_6=U\,X^{-1}$ are {\it basic} decorated Polyakov loops and 
 $L_7=X\,V\,U,\, L_8=X^{-1} U^2,\, L_9 = X\,V^2,\, L_10=X^{-1} U\,V ,\,
 L_{11} = L_{12}= U\,V$ 
 are {\it primary} decorated Polyakov loops obtained by chaining a basic
 straight Polyakov loop and a basic decorated Polyakov loop. Notice that we can
 attach further straight Polyakov loops and obtain primary loops
 of arbitrary length in order $\kappa^2$
 but we stopped at length $l=6$ for this illustration.
  
Since the lattice is even we take only even power of $Q$ in 
evaluating Eq. (\ref{e.dloop2}). We shall only consider loops up to length 6.  
To order 0 in
$\kappa$ we find:
\bear
\frac{1}{2n} Q^{2 n}= \frac{1}{n}\left(4 \zu \epsilon U\right)^n +
\frac{1}{n}\left(4 \zd \epsilon U^{-1}\right)^n 
+\frac{1}{n}\left(4 \zu \epsilon V\right)^n +\frac{1}{n}
\left(4 \zd \epsilon V^{-1}\right)^n \label{e.exphdm} 
\eear
Notice that these basic loops are of lengths 2 ,
$U =U_1U_2,\, V=V_1V_2$. 
The calculation was performed up to order 16 , which is $2 n_{max}$ 
and one can check that up to this order
there is no deviation from the logarithm series. 
Assuming  that
the series continues as expected we find to this order
\bear
D^{[0]} &=& (1-4\epsilon \zu^2  U) (1-4\epsilon \zu^2  V) 
(1-4\epsilon \zd^2 U^{-1}) (1-4\epsilon \zd^2  V^{-1}) \nn \\
&=& 1-4\epsilon \zu^2  (U + V) +16 \zu^4 U\,V + O(\kappa^4) \label{hd2_0} 
\eear
A similar result follows from the  
decorated Polyakov loops which appear with a factor $\kappa^2$ 
(for simplicity we no longer write the non-dominant
inverse loops which have
factors $\zd$):
\bear
D^{[2]}_1&=&
(1-4\epsilon \zu^2 \kappa^2 {U\,X^{-1}}) 
(1-4\epsilon \zu^2  \kappa^2 {V\,X}) 
\nn \\
&=& 1 
-4 \epsilon \kappa^2 \zu^2 (V\,X
+ U\,X^{-1}) +O(\kappa^4) \label{hd2_2_1}
\eear
Finaly the non-basic primary loops $L_7 - L_{12}$ give
\bear
D^{[2]}_{2}&=&
(1-16 \zu^4  \kappa^2 X\,V\,U) 
(1-16 \zu^4 \kappa^2 X^{-1}\,U^2)\nn \\ 
&&\times (1-16 \zu^4 \kappa^2 X\,V^2 ) 
(1-16 \zu^4  \kappa^2 X^{-1}U\,V)\nn \\
&&\times (1-16 \zu^4  \kappa^2 U\,V)^2 \nn
\\
&=& 1 
-16 \kappa^2 \zu^4 (2\,U\,V + X\,V\,U + X^{-1}U\,V +X^{-1}U^2 +X\,V^2))
\nn\\&&+O(\kappa^4) 
\eear

We obtain thus the HD determinant to order $\kappa^2$
including all loops up to length 6, 
in complete agreement with the exact determinant to this order
\bear
D^{[2]} &=& 
\left(1-4\epsilon \zu^2  (U + V) +16 \zu^4 U\,V\right)
\left(1-4\epsilon \zu^2 \kappa^2  ( X\,V+ X^{-1}U)\right) \nn \\
&&\times \left(1-16\zu^4 \kappa^2 (U\,V (2 + X+ X^{-1}) +X^{-1}U^2 +X\,V^2) \right) 
\nn \\
 &=&1 - 4\epsilon \zu^2 (U+V) + 16 \zu^4 U\,V \nn \\
&&- 4\epsilon \zu^2 \kappa^2 (V\,X+U\,X^{-1}) -32  \zu^4 \kappa^2 U\,V  \\
D^{[2]}_{\rm exact} &=& 
1 - 4\epsilon \zu^2 (U+V) + 16 \zu^4 U\,V \nn \\
&&- 4\epsilon \zu^2 \kappa^2 (V\,X+U\,X^{-1}) -32  \zu^4 \kappa^2 U\,V .
\eear


\newpage
\begin{thebibliography}{99}

\bibitem{Heisenberg:1935qt}
  W.~Heisenberg and H.~Euler,
  Z.\ Phys.\  {\bf 98} (1936) 714
  [physics/0605038].

\bibitem{Schwinger:1951nm}
  J.~S.~Schwinger,
  Phys.\ Rev.\  {\bf 82} (1951) 664.

\bibitem{seiler} E. Seiler, in: Proceedings of the International Summer
School of Theoretical Physics, Poiana Brasov, Romania, 1981, edited by P.
Dita, V. Georgescu, and R. Purice, Progress in Physics
(Birkh\"auser, Boston, 1982), Vol. 5, p. 263-310.

\bibitem{Haba:1981rp}
  Z.~Haba,
  Phys.\ Rev.\ D {\bf 26} (1982) 3506;
  Z.~Haba,
  Phys.\ Rev.\ D {\bf 29} (1984) 1718.

\bibitem{Langfeld:2002vy}
  K.~Langfeld, L.~Moyaerts and H.~Gies,
  Nucl.\ Phys.\ B {\bf 646} (2002) 158
  [hep-th/0205304].

  
\bibitem{Schmidt:2003bf}
  M.~G.~Schmidt and I.~Stamatescu,
  Mod.\ Phys.\ Lett.\ A {\bf 18} (2003) 1499.


\bibitem{Fry:2015qua}
  M.~P.~Fry,
  Phys.\ Rev.\ D {\bf 91} (2015) 085026
  [arXiv:1504.03117 [hep-th]].

\bibitem{Bender:1992gn}
  I.~Bender, T.~Hashimoto, F.~Karsch, V.~Linke, A.~Nakamura, M.~Plewnia, I.~O.~Stamatescu and W.~Wetzel,
  Nucl.\ Phys.\ Proc.\ Suppl.\  {\bf 26} (1992) 323.

\bibitem{Blum:1996xt}
  T.~Blum, J.~E.~Hetrick and D.~Toussaint,
  Nucl.\ Phys.\ Proc.\ Suppl.\  {\bf 47} (1996) 543.

\bibitem{Engels:1999tz}
  J.~Engels, O.~Kaczmarek, F.~Karsch and E.~Laermann,
  Nucl.\ Phys.\ B {\bf 558} (1999) 307
  [hep-lat/9903030].

\bibitem{Aarts:2001dz}
  G.~Aarts, O.~Kaczmarek, F.~Karsch and I.~O.~Stamatescu,
  Nucl.\ Phys.\ Proc.\ Suppl.\  {\bf 106} (2002) 456
  [hep-lat/0110145].

\bibitem{dfss}
  R.~De Pietri, A.~Feo, E.~Seiler and I.~O.~Stamatescu,
  Phys.\ Rev.\ D {\bf 76} (2007) 114501
  [arXiv:0705.3420 [hep-lat]].

\bibitem{Fromm:2012uj}
  M.~Fromm, J.~Langelage, S.~Lottini, M.~Neuman and O.~Philipsen,
  J.\ Phys.\ Conf.\ Ser.\  {\bf 432} (2013) 012033
  [arXiv:1210.7994 [hep-lat]].

\bibitem{Aarts:2015yba}
  G.~Aarts, F.~Attanasio, B.~J\"ager, E.~Seiler, D.~Sexty and 
  I.~O.~Stamatescu,
  Acta Phys.\ Polon.\ Supp.\  {\bf 8} (2015) 2,  405
  [arXiv:1506.02547 [hep-lat]].

\bibitem{Stamatescu:1980br}
  I.~O.~Stamatescu,
  Phys.\ Rev.\ D {\bf 25} (1982) 1130.

\end{thebibliography}
\end{document}